\documentclass[reprint,twocolumn,aps,prb,superscriptaddress,floatfix]{revtex4-2}

\usepackage{amsmath}    
\usepackage{amssymb}    
\usepackage{graphicx}   
\usepackage{dcolumn}    
\usepackage{tabularx}   
\usepackage{bm}         
\usepackage{longtable}  
\usepackage{multirow}   
\usepackage{float}      
\usepackage{natbib}
\usepackage{color}

\begin{document}
	\title{Muon spin rotation and relaxation study on topological noncentrosymmetric superconductor PbTaSe$_2$}
	
	\author{Z.~H.~Zhu}
	\author{C.~Tan}
	\author{J.~Zhang}
	\affiliation{State Key Laboratory of Surface Physics, Department of Physics, Fudan University, Shanghai 200433, China}	
	
	\author{P.~K.~Biswas}
	\affiliation{Laboratory for Muon Spin Spectroscopy, Paul Scherrer Institut, CH-5232 Villigen PSI, Switzerland}
	\affiliation{ISIS Pulsed Neutron and Muon Source, STFC Rutherford Appleton Laboratory, Harwell Campus, Didcot, Oxfordshire OX11 0QX, United Kingdom}
	\author{A.~D.~Hillier}
	\affiliation{ISIS Pulsed Neutron and Muon Source, STFC Rutherford Appleton Laboratory, Harwell Campus, Didcot, Oxfordshire OX11 0QX, United Kingdom}
	\author{M.~X.~Wang}
	\author{Y.~X.~Yang}
	\author{C.~S.~Chen}
	\author{Z.~F.~Ding}
	\affiliation{State Key Laboratory of Surface Physics, Department of Physics, Fudan University, Shanghai 200433, China}
	
	\author{S.~Y.~Li}
	\author{L.~Shu}
	\affiliation{State Key Laboratory of Surface Physics, Department of Physics, Fudan University, Shanghai 200433, China}
	\affiliation{Collaborative Innovation Center of Advanced Microstructures, Nanjing 210093, China}
	\affiliation{Shanghai Research Center for Quantum Sciences, Shanghai 201315, China}
	
	\begin{abstract}
		
		Topological superconductivity is an exotic phenomenon due to the symmetry-protected topological surface state, in which a quantum system has an energy gap in the bulk but supports gapless excitations confined to its boundary. Symmetries including central and time-reversal (TRS), along with their relations with topology, are crucial for topological superconductivity. We report muon spin relaxation/rotation ($\mu$SR) experiments on a topological noncentrosymmetric superconductor PbTaSe$_2$ to study its TRS and gap symmetry. Zero-field $\mu$SR experiments indicate the absence of internal magnetic field in the superconducting state, consistent with previous $\mu$SR results. Furthermore, transverse-field $\mu$SR measurements reveals that the superconducting gap of PbTaSe$_2$ is an isotropic three-dimensional fully-gapped single-band. The fully-gapped results can help understand the pairing mechanism and further classify the topological superconductivity in this system.
	
	\end{abstract}

	
	\maketitle
	
	\section{Introduction}
	
	In unconventional superconductors, symmetries in addition to U(1) gauge symmetry are broken in the superconducting state, leading to exotic and potentially useful properties, therefore realization and study of superconductivity in systems with reduced symmetry is one of the most crucial research field.  Among these, noncentrosymmetric crystal structures with significant spin-orbital coupling are of particular interest \cite{Bauer2012}. In superconductors with noncentrosymmetric crystal structures, the absence of inversion symmetry leads to the splitting of the Fermi surfaces into two opposite spin configurations, and results in the mixed singlet-triplet nature in the order parameter \cite{Bauer2012,Bauer2004,Frigeri2004,Gorkov2001}. As a result, it can give rise to a range of novel phenomena, including the recently proposed topological superconductivity \cite{Alicea2012,Kim2018,Scheurer2015,Sun2015}.
	
	Topological superconductivity is an exotic phenomenon due to the symmetry-protected topological surface state~\cite{Qi2011}. In a topological superconductor, the bulk state is a fully gapped superconducting state, while the surface state is a metal state. The Hamiltonian of such state is defined by several important symmetries. The most important symmetry for a topological material is time-reversal symmetry (TRS), which determines the topological mode of the material \cite{Qi2011}. TRS is also one of the most intensely studied symmetries for a superconductor, and has been observed in a handful of weakly correlated noncentrosymmetric superconductors \cite{Hillier2009,Biswas2013,Singh2014,Barker2015,Singh2017,Singh2018}. Despite broken TRS being a clear signature of unconventional superconductivity was observed, many other properties resemble conventional superconductors. This immediately raises an important question, namely what is the origin of the TRS breaking in this kind of material, or does TRS breaking occur together with a conventional electron-phonon pairing mechanism? Furthermore, the relationship between TRS breaking and breaking inversion symmetry requires clarification since there are also examples in which TRS breaking occurs in centrosymmetric systems \cite{Hillier2012}.
			
	Recently, a noncentrosymmetric superconductor PbTaSe$_2$ with transition temperature $T_c=3.7$~K was reported to host a $\mathbb{Z}_2$ topological state with topological nodal-line state by \textit{ab initio} calculations, angle-resolved photoemission spectroscopy (ARPES) and soft point-contact spectroscopy experiments \cite{Ali2014,Bian2016,Chang2016,Guan2016,Chen2016,Le2020}. Zero-field muon spin relaxation ($\mu$SR) measurement show no evidence for a TRS breaking field greater than 0.05 G in the superconducting state~\cite{Wilson2017}. Different techniques, including specific heat and nuclear magnetic resonance (NMR) measurements~\cite{Ali2014,Maeda2018}, agree with an in-plane fully gapped superconducting state of PbTaSe$_2$. A recent calculation work suggest multi-band superconductivity due to the complex band structure revealed by ARPES~\cite{Lian2019,Bian2016,Chang2016}.  However, experimentally it remains controversial that whether it is single band or multi-band superconductor. The tunnel diode oscillator (TDO) experiments support single band~\cite{Pang2016}, but the thermal conductivity measurements suggest multi-band picture \cite{Wang2016}. In addition, while both scanning tunnel microscopy (STM) and $\mu$SR results can be described by either single or multi-band model, STM results give similar magnitude of gap values from different band, and $\mu$SR results indicate two different gaps \cite{Wilson2017,Guan2016}. 
	
	It is worth noting that PbTaSe$_2$ is a 3-dimensional material with strong anisotropic behavior~\cite{Zhang2016}. All the superconducting pairing symmetry studies of PbTaSe$_2$ were in $ab$-plane so far, due to the limitation of measurement along the $c$-direction. It is particularly important to perform the gap symmetry study along $c$-direction, to clarify the relationship between topology, TRS, and superconducting pairing symmetry of this noncentrosymmetric superconductor.
	
	We report the $\mu$SR experiment results on single crystalline PbTaSe$_2$.  No evidence of TRS breaking is confirmed. We find fully-gapped superconductivity in both in-plane and out-of-plane directions. Our results prefer the single-band picture. Most intriguingly, the normalized superfluid density in two directions have exactly same temperature dependence, suggesting a possible isotropic three-dimensional gap. 
	
	\section{Experimental Details}
	
	PbTaSe$_2$ single crystals were grown by the chemical vapor transport (CVT) method as previously reported \cite{Bian2016}. The typical size of obtained single crystals is $5\times5\times0.02$~mm$^3$. The quality of the single crystals was checked by X-ray diffraction (XRD), magnetic susceptibility and resistivity measurements \cite{Wang2016}. 
	
	$\mu$SR experiments were performed at the DOLLY beam line at Paul Scherrer Institut, Villigen, Switzerland. A mosaic of single crystals were stacked and aligned with the (001) basal plane attached to a copper sample holder using dilute GE varnish. Helium-3 cryostat was used to cool the sample down to 0.25~K. In a $\mu$SR experiment, spin-polarized positive muons are implanted into a sample. On decay of the muon after an average lifetime of 2.2~$\mu$s, a positron is emitted preferentially along the direction of the muon spin. The time evolution of muon spin polarization is determined by detecting decay positrons from an ensemble of 1-2 $\times$ 10$^7$ muons. The functional form of the muon spin polarization depends on the spatial distribution and dynamical fluctuations of the muon magnetic environment.
	
	During the experiments, the initial muon spin is 45$^\circ$ from the $c$-axis, which was surrounded by four detectors: Forward, Backward, Up, and Down. Hence we can measure the muon spin polarization along two different directions, i.e. parallel and perpendicular to the $c$-axis. As a trade, due to the angle between muon spin and detectors, the initial asymmetry in our experiments is much lower than the common value that is about 0.25. Zero-field (ZF) $\mu$SR was performed above and below $T_c$ to study whether there is spontaneous small magnetic field in the superconducting state due to the TRS breaking \cite{Luke1998,Aoki2003,Hillier2009}. 
	
	In transverse-field (TF) $\mu$SR experiments, an external magnetic field $\mu_0H$ (field cooled from above $T_c$ in a superconductor) was applied to induce a flux-line lattice (FLL) where the internal magnetic field distribution is determined by the magnetic penetration depth $\lambda$, the vortex core radius and the structure of the FLL. The external field $\mu_0H$ should be between $\mu_0H_{c1}$ and $\mu_0H_{c2}$. In our case, $\mu_0H_{c1}$ is about 4-9~mT \cite{Ali2014,Zhang2016,Wilson2017}, $\mu_0H_{c2}$ along $c$-axis is about 0.32~T, and $\mu_0H_{c2}$ parallel to $ab$-plane is about 1.25~T \cite{Zhang2016}. 
	
	The muon spin relaxation rate is related to the root-mean-square (rms) width of the internal magnetic field distribution in the FLL, and hence also related to $\lambda$, the details of which will be discussed after presenting the experimental results. We first applied magnetic field parallel to $c$-axis to obtain the penetration depth in the $ab$-plane and study the gap symmetry in that direction \cite{Brandt2003}, and compare with the published work \cite{Wilson2017}. Then we apply an external field normal to $c$-axis. Since the samples were not aligned along $a$ or $b$-axes, the gap symmetry in $ac$ or $bc$-plane cannot be obtained. Instead, we obtained the average gap symmetry in planes that includes $c$-axis but with random directions in $ab$-plane. If there is any node in the superconducting gap out of $ab$-plane, the temperature dependence of superfluid density should deviate from the $s$-wave behavior. 
	
	The $\mu$SR data were analyzed with \textsc{musrfit} software package \cite{Suter2012}. 
	
	\section{Results}
	
		\subsection{ZF-$\mu$SR} 
		
		\begin{figure}
			\includegraphics[clip,width=8cm]{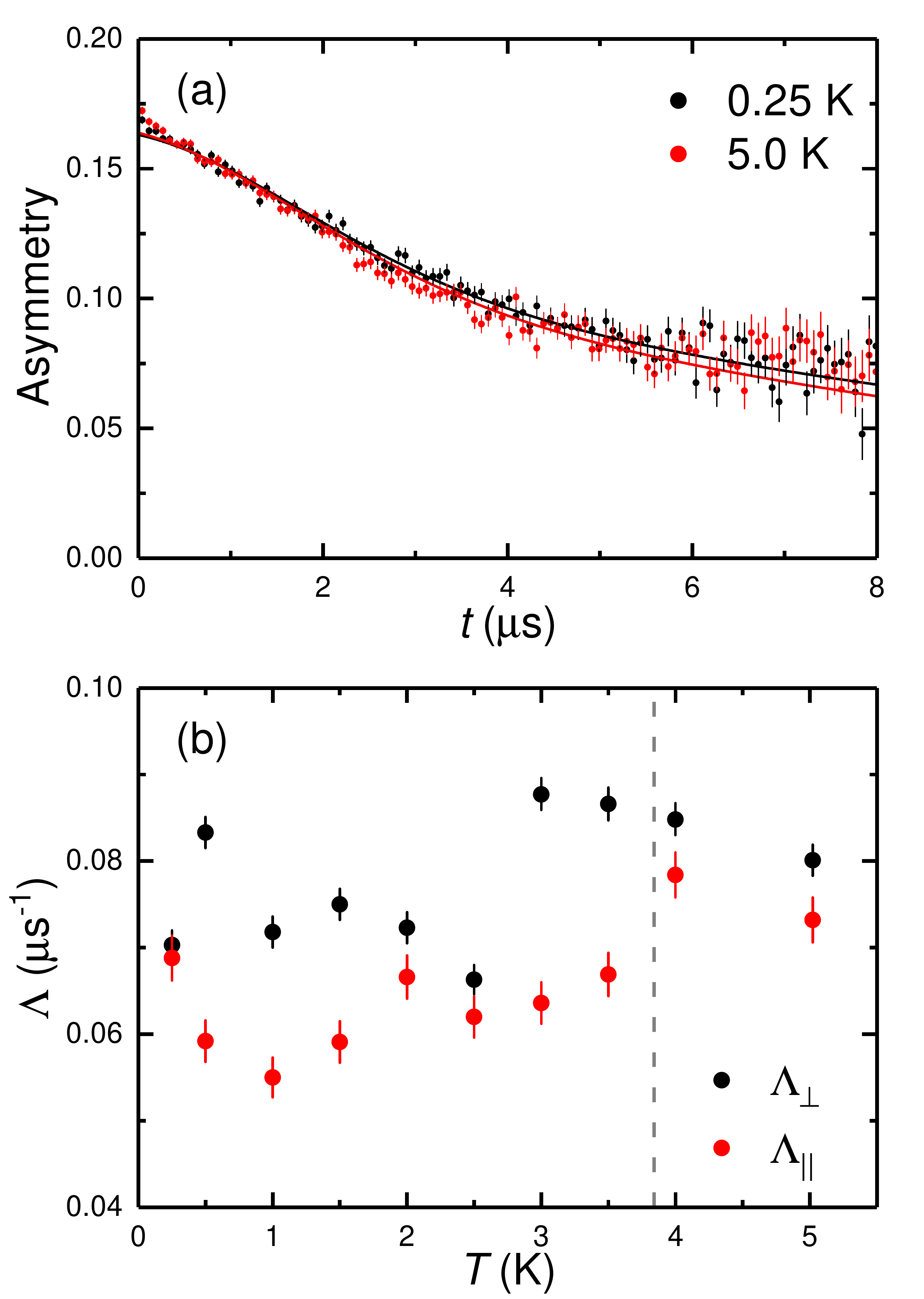}
			\caption{Zero-field $\mu$SR. (a) $\mu$SR asymmetry spectra $\mathrm{asy}(t)$. Black  circles: superconducting state. Red  circles: normal state. Solid curves: fits to the data with Eq. \ref{ZF-FittingFunction}. (b) Temperature dependence of relaxation rate $\Lambda$. Dashed line marks $T_c$. }
			\label{Fig1}
		\end{figure}
		
		Representative ZF asymmetry time spectra at selective temperatures are shown in Fig. \ref{Fig1}(a). No significant difference can be observed between the data above and below $T_c=3.84$~K. The $\mu$SR asymmetry spectrum consists of two contributions: a signal from muons stop in the sample and a slowly relaxing background signal from muons that stop in the copper sample holder. The spectra in both directions can be well described by the function
		\begin{align}
		\mathrm{asy}(t)=a_0[f\exp(-\Lambda t)+(1-f)G\mathrm{_{KT}^{dyn}}(\sigma_\mathrm{Cu}^\mathrm{ZF},\nu,t)],
		\label{ZF-FittingFunction}
		\end{align}
		where the first and second terms represent sample and background signals, respectively. Here $a_0$ is the initial asymmetry, and $f $ denotes the fraction of muons stopping in the sample. The data of the first 0.1~$\mu$s is dropped to avoid the early-time problems. The temperature independent $f=0.68$ is determined from TF-$\mu$SR. The dynamic ZF Kubo-Toyabe (KT) function $G\mathrm{_{KT}^{dyn}}$, which was used previously to fit ZF-$\mu$SR data of Cu~\cite{Hayano1979,Kadono1989,Clawson1983}, describes the data adequately. We obtain $\sigma_\mathrm{Cu}^\mathrm{ZF} = 0.38$ $\mu$s$^{-1}$ and $\nu = 0.4$~MHz, same as previously reported \cite{Kadono1989,Clawson1983}. 
		
		The temperature dependence of the ZF relaxation rate $\Lambda$ is shown in Fig.~\ref{Fig1}(b). Consistent with previous report~\cite{Wilson2017}, no significant change crossing $T_c$ is observed down to 0.25~K in our study. Such results suggest that there is no spontaneous magnetic field appearing in the superconducting state. Therefore, there is no TRS breaking, excluding the existence of triplet pairing~\cite{Wilson2017}. Recently, it is reported that muon could modify its local environment in ZF $\mu$SR, making $\mu$SR results on TRS differ from other techniques \cite{Huddart2021}. This is not in our case since our result is consistent with previous $\mu$SR work, and there is no results using other techniques to study TRS so far. 
		
		\subsection{TF-$\mu$SR}
		
			\subsubsection{$H \parallel c$-axis}
			
\begin{figure}
				\includegraphics[clip,width=8cm]{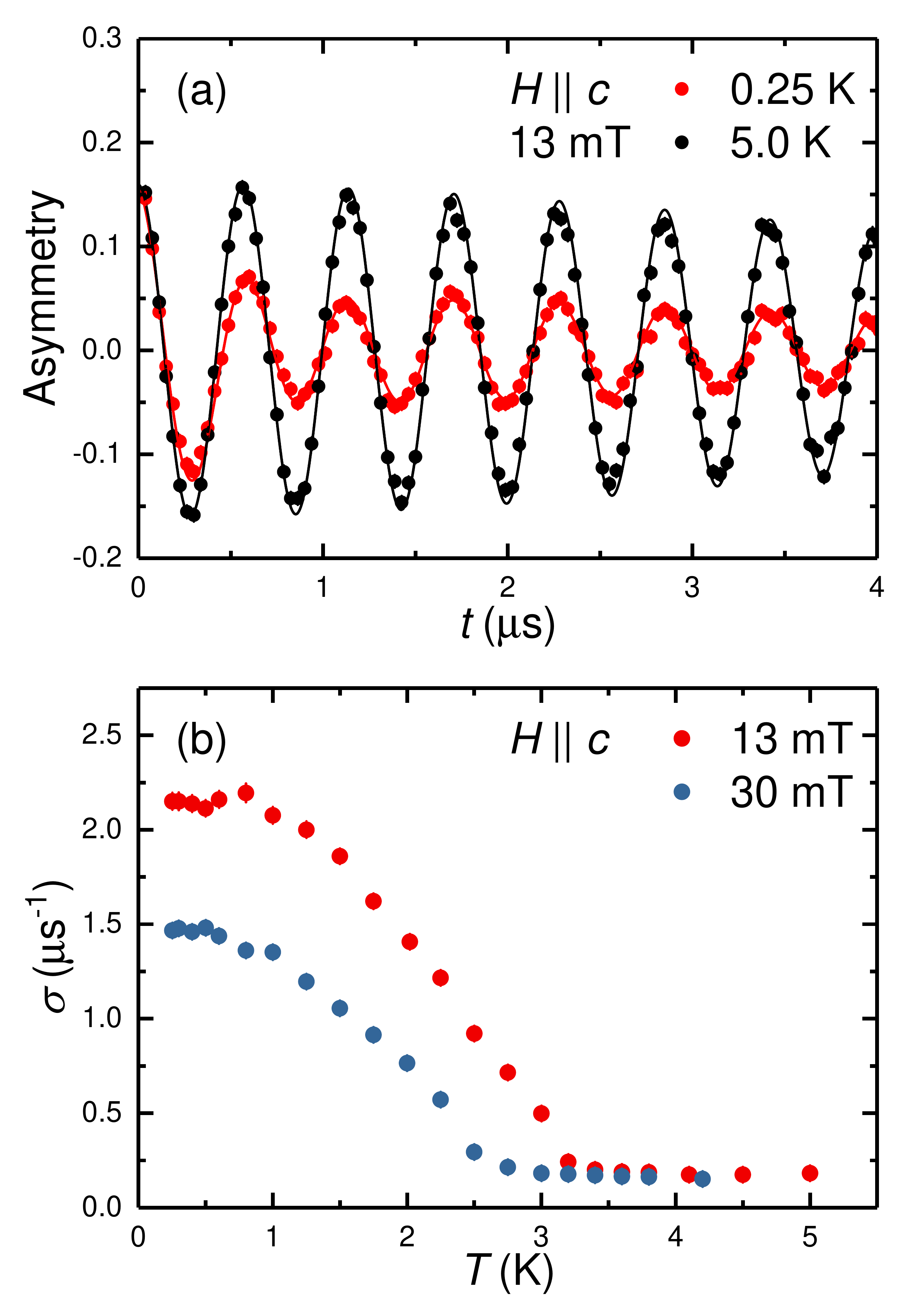}
				\caption{Transverse-field $\mu$SR with external field $H$ parallel to $c$-axis. (a) $\mu$SR asymmetry spectra $\mathrm{asy}(t)$, $\mu_0H=13$~mT. Black  circles: normal state. Red  circles: superconducting state. Solid curves: fits to the data with Eq. \ref{TF-FittingFunction}. (b) Temperature dependence of relaxation rate $\sigma$ measured in 13 and 30~mT. }
				\label{Fig2}
			\end{figure}

			Fig.~\ref{Fig2}(a) shows the TF-$\mu$SR muon spin precession signals at applied field of 13~mT in the normal and superconducting states of PbTaSe$_2$. As seen in Fig.~\ref{Fig2}(a), in the superconducting state the damping of signal is enhanced due to the field broadening generated by the vortex lattice. 
			
			The TF-$\mu$SR asymmetry spectra in PbTaSe$_2$ can be well described by the function
			\begin{align}
			\mathrm{asy}(t)&=a_0\{f\exp[-\frac{(\sigma t)^2}{2}]\cos(\gamma_\mu\mu_0H_\mathrm{int} t+\varphi)\notag\\
			&+(1-f)\exp[-\frac{(\sigma_\mathrm{Cu}^\mathrm{TF} t)^2}{2}]\cos(\gamma_\mu\mu_0H t+\varphi)\},
			\label{TF-FittingFunction}
			\end{align}
			where the first and second terms represent sample and background signals, respectively. The relaxation rate of copper $\sigma_\mathrm{Cu}^\mathrm{TF} = 0.24$~$\mu$s$^{-1}$ is also temperature independent, consistent with previous report \cite{Camani1977}. The ratio of ZF and TF relaxation rate of copper is 1.58, consistent with the theoretical value~\cite{Hayano1979}. The Gaussian relaxation rate $\sigma$ from the sample is due to nuclear dipolar fields in the normal state and it enhanced in the superconducting state by the vortex lattice. $\gamma_\mu/2\pi=135.5$~MHz/T is the gyromagnetic ratio of muon, $H_\mathrm{int}$ is the internal field, which is reduced due to diamagnetic screening. The curves in Fig.~\ref{Fig2}(a) are the fits of Eq.~(\ref{TF-FittingFunction}).
			
			Temperature dependence of $\sigma$ is shown in Fig. \ref{Fig2}(b) at two different applied magnetic fields. The temperature independence of $\sigma$ above $T_c$ and the increase of $\sigma$ with decreasing temperature below $T_c$ are observed, indicating the bulk superconductivity occurs below $T_c$.  The lower $T_c$ in the 30~mT is consistent with the suppressing effect on superconductivity by external magnetic field. 
			
			\begin{figure}
				\includegraphics[clip,width=8cm]{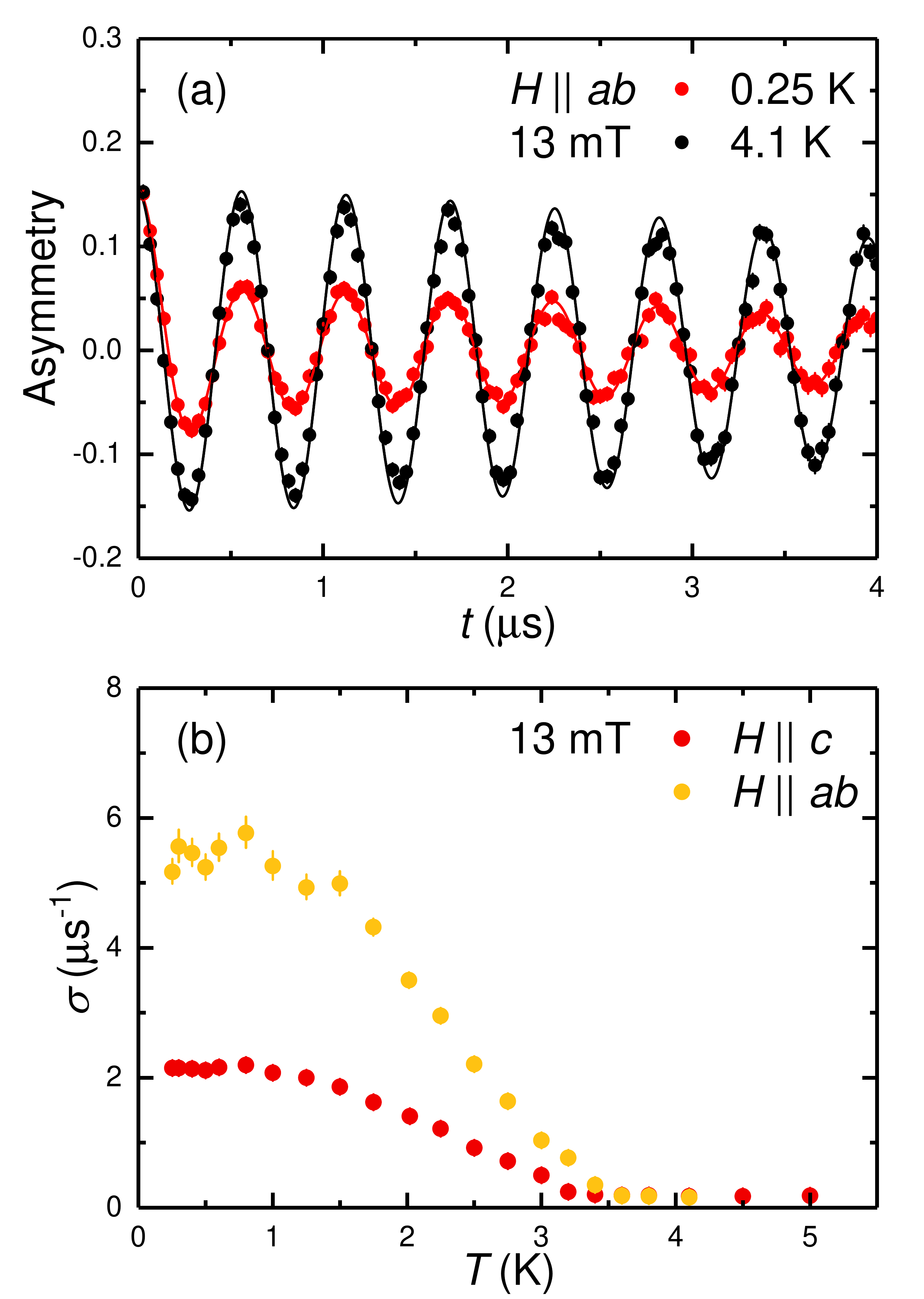}
				\caption{Transverse-field $\mu$SR with external field $\mu_0H=13$~mT parallel to $ab$-plane. (a) $\mu$SR asymmetry spectra $\mathrm{asy}(t)$. Black  circles: normal state. Red  circles: superconducting state. Solid curves: fits to the data with Eq. \ref{TF-FittingFunction}. (b) Temperature dependence of relaxation rate $\sigma$ measured in two directions. Red circles: $\mu_0H \parallel c$. Yellow circles: $\mu_0H \parallel ab$. }
				\label{Fig3}
			\end{figure}
		
			\subsubsection{$H \parallel ab$-plane}
		
			Similar results were obtained with field parallel to $ab$-plane, shown in Fig.~\ref{Fig3}. The $\mu$SR asymmetry spectra can also be well described by Eq.~(\ref{TF-FittingFunction}). However, compared with $\sigma$ in 13~mT field along $c$-axis, the relaxation rate $\sigma$ here is much larger, suggesting a much broader field distribution. Similar results were also reported in Mo$_3$P \cite{Shang2019}. Such difference between different directions can be attributed to the strong anisotropy of $H_{c2}$ \cite{Brandt2003}. The estimated $\mu_0H_{c2}$ along $c$-axis is 0.32~T, while $\mu_0H_{c2}$ is 1.25~T parallel to $ab$-plane determined from electrical resistivity measurements~\cite{Zhang2016}. In Fig.~\ref{Fig3}(b), the larger $T_c$ measured at magnetic field parallel to $ab$-plane also indicate larger $H_{c2}$ in that direction. 
			
			\subsubsection{Pairing Symmetry}
			
			The Gaussian relaxation rate $\sigma$ is related to the Gaussian internal field distribution \cite{Hayano1979}. For a type-II superconductor in vortex state, the internal field distribution is convolution of contribution from the vortex lattice and nuclear dipole field distribution of the host material. Thus $\sigma$ is given by
			\begin{align}
			\sigma^2=\sigma_\mathrm{SC}^2+\sigma_\mathrm{dip}^2,
			\end{align}
			where $\sigma_\mathrm{SC}$ is the vortex lattice contribution, and $\sigma_\mathrm{dip}$ is temperature independent in the normal state and is not expected to change in the superconducting state. After determining $\sigma_\mathrm{dip}$ = 0.181(7)~$\mu$s$^{-1}$ from the normal state data, we can get temperature dependence of $\sigma_\mathrm{SC}$. 
			
			\begin{figure}
				\includegraphics[clip,width=8cm]{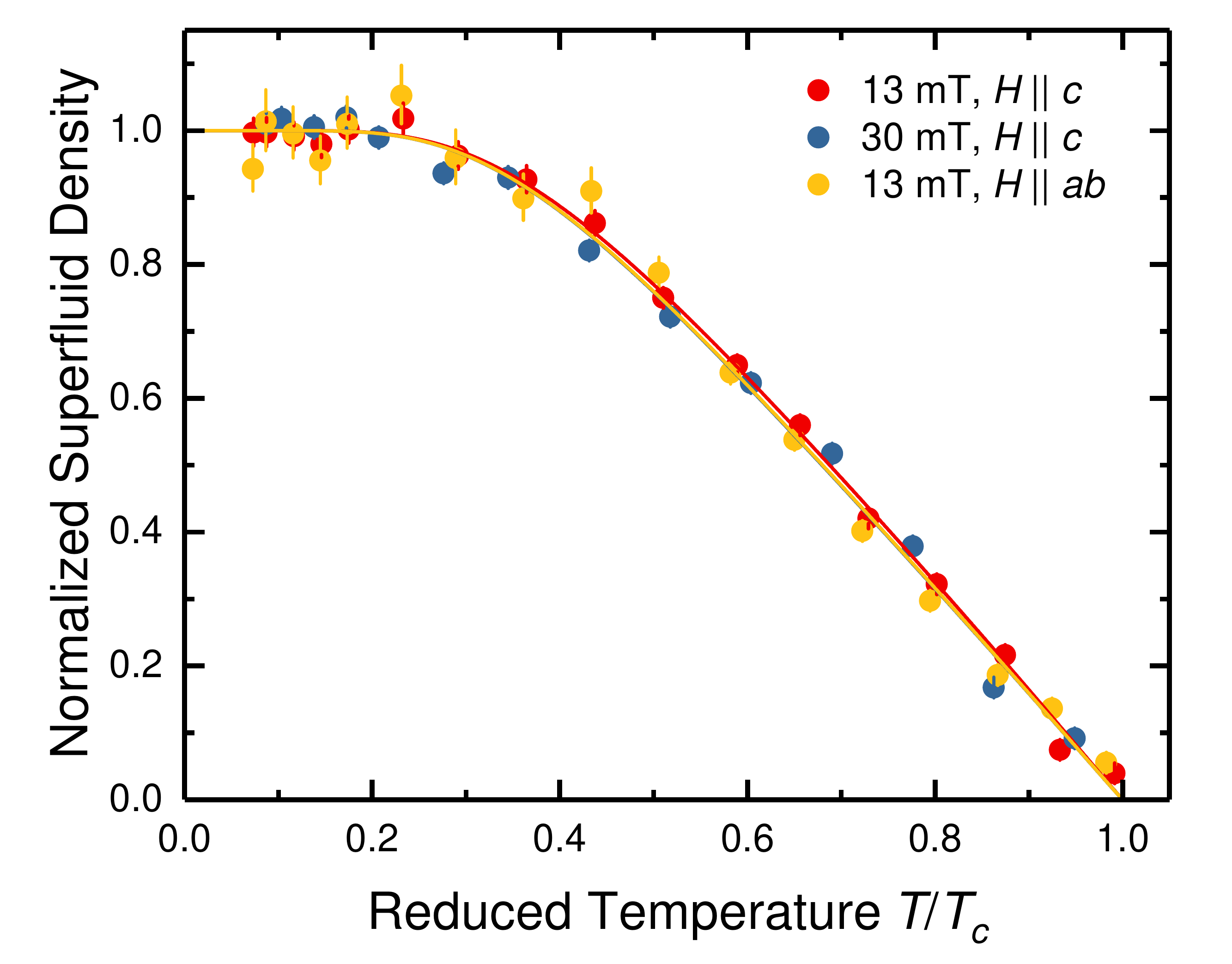}
				\caption{Normalized superfluid density $n_s(T)/n_s(0)$ plotted versus reduced temperature $T/T_c$. Red  circles: $\mu_0H=13$~mT parallel to $c$-axis. Blue  circles: $\mu_0H=30$~mT parallel to $c$-axis. Yellow  circles: $\mu_0H=13$~mT parallel to $ab$-plane. Solid curves: fits to the data with Eq. \ref{GapFit}.}
				\label{Fig4}
			\end{figure}
			
			On the other hand, for a type-II superconductor, the internal field distribution can be described by penetration depth $\lambda$, which can be estimated based on $\mu_0H_{c1}=9$~mT \cite{Zhang2016,Brandt2003}:			
			\begin{align}
			\mu_0H_{c1}=\frac{\Phi_0}{4\pi\lambda^2}(\ln\frac{\lambda}{\xi}+0.497),
			\end{align}
			where $\Phi_0=2.07\times10^{-15}$~Wb is the fluxoid quantum, and $\xi$ is the coherence length. Based on the well-known relation
			\begin{align}
			\mu_0H_{c2}=\frac{\Phi_0}{2\pi\xi^2},
			\end{align}
			we obtain $\xi_c=32.1$~nm and $\xi_{ab}=16.2$~nm, where $\xi_c$ and $\xi_{ab}$ are the coherence length parallel to $c$-axis and $ab$-plane, respectively \cite{Zhang2016}. Then we can estimate the value of $\lambda$ along $c$-axis $\lambda_c=208.1$~nm and in $ab$-plane $\lambda_{ab}=242.0$~nm, and Ginzburg-Landau parameter $\kappa=\lambda/\xi$ can be further derived, which shows $\kappa_c=6.5$ and $\kappa_{ab}=14.9$. With $\kappa>5$, and a not-too-small reduced magnetic field $h=H/H_{c2}>0.25/\kappa^{1.3}$, one can calculate penetration depth $\lambda$ more accurately using \cite{Brandt2003}
			\begin{align}
			\sigma_\mathrm{SC}=0.172\frac{\gamma_\mu\Phi_0}{2\pi}(1-h)[1+1.21(1-\sqrt{h})^3]\lambda^{-2}.
			\end{align}
			That is to say, $\sigma_\mathrm{SC}$ is proportional to $\lambda^{-2}$, and the complicated coefficient can be reduced by normalizing $\sigma_\mathrm{SC}(T)$ to $\sigma_\mathrm{SC}(0)$.  
						
			Based on the London approximation, the superfluid density $n_s(T)$ is also proportional to $\lambda^{-2}$. For a fully gapped $s$-wave superconductor, $n_s$ can also be written as
			\begin{align}
			\frac{\sigma_\mathrm{SC}(T)}{\sigma_\mathrm{SC}(0)}=\frac{n_s(T)}{n_s(0)}=1+2\int_\Delta^\infty\frac{\partial f}{\partial E} \frac{E}{\sqrt{E^2-\Delta^2}}dE,
			\label{GapFit}
			\end{align}
			where $n_0$ is the superfluid density at zero temperature, $E$ is the energy difference above the Fermi energy, $f=1/[\exp(E/k_BT)+1]$ is the Fermi function, $k_B=8.617\times10^{-5}$~eV/K is the Boltzmann's constant, and $\Delta$ is the gap function. For a fully-gapped $s$-wave superconductor, the temperature dependence of $\Delta$ can be approximated by
			\begin{align}
			\Delta(T)=\Delta_0\tanh\{1.82[1.018(\frac{T_c}{T}-1)]^{0.51}\},
			\end{align}
			where $\Delta_0$ is the zero temperature gap \cite{Carrington2003}. 
		
			The fitting results of normalized superfluid density $n_s(T)/n_s(0)$ are plotted in Fig. \ref{Fig4}. All three groups of data can be well fitted by single gap $s$-wave model. The derived zero temperature gap $\Delta_0$ is 0.463(7)~meV for $\mu_0H=13$~mT along $c$-axis, 0.383(11)~meV for $\mu_0H=30$~mT along $c$-axis, and 0.458(15)~meV for $\mu_0H=13$~mT parallel to $ab$-plane. Interestingly, all three curves stack together and share the same behavior. Besides, unlike other anisotropic properties, even including the relaxation rates $\sigma_\mathrm{SC}$ for the superfluid density fitting, the two gap derived from two directions are very close (0.463(7) and 0.458(15)~meV). 
			
	\section{Discussion}	
					
		Previous work using $\mu$SR reported that two-gap model could describe the in-plane behavior of PbTaSe$_2$ better \cite{Wilson2017}. The evidence is not strong enough since there is no critical difference, and there is only one slightly increased point that influenced the conclusion. Our experiments can only be performed down to 0.25~K, just missing the critical point. Based on our results, PbTaSe$_2$ is a fully gapped superconductor, and whether it is single gap or two gap needs experiments to a lower temperature with dilution refrigerator. 	

		Comparing our experimental results with previous theoretical calculations, we can find similar isotropic superconducting gap around $H$ point in reciprocal space as defined in Ref.~25. The magnitude of gap derived from our results is also consistent with former reports \cite{Wilson2017,Guan2016}, but slightly smaller than the calculated values~\cite{Lian2019}. It could come from the suppression effect of external fields, which can be seen from the suppressed gap in our results. Although there are several other superconducting gaps derived by the theory, none of them is dominant in our experimental results. However, such a complicated band structure could account for the multi-band experimental results. 
		
		The anisotropy of relaxation rate in different directions suggest an anisotropic penetration depth $\lambda$. Since $\lambda$ relates to effective mass, the strong anisotropy suggests a possible tensor effective mass. This point is further supported by the anisotropic $\mu_0H_{c2}$ and coherence length $\xi$. Noticed that spin-orbital coupling plays the most significant role at $H$ point that induces topological properties~\cite{Guan2016,Bian2016}, such complex phenomenon is easy to expect. 
		
		Given that the dominant superconducting gap of PbTaSe$_2$ is the isotropic gap around $H$ point, PbTaSe$_2$ is a 3D material. Besides, the fully gapped picture is also consistent with topological superconductivity. 
		The presence of TRS is consistent with the picture of a 3D $\mathbb{Z}_2$ topological superconductor \cite{Chen2016,Chang2016,Schnyder2008}. Based on the classification of topological superconductivity \cite{Kitaev2009,Schnyder2008}, besides TRS, particle-hole symmetry (PHS) and chiral symmetry (SLS) are also symmetries of great significance. It is important to study PHS and SLS of PbTaSe$_2$ to understand its topological properties better. Furthermore, it would be more intriguing to study the role of the absence of inversion symmetry in its topological properties. 

	\section{Conclusion}
	
		In summary, we performed ZF and TF-$\mu$SR experiments on single crystalline PbTaSe$_2$. The preservation of TRS is confirmed by ZF-$\mu$SR. The pairing symmetry is derived from TF-$\mu$SR, indicating an isotropic 3D fully-gapped single-band picture, satisfying the requirement of topological superconductivity. The complicated band structure could account for the multi-band picture, but the other bands are not dominant in superconductivity. 

	\begin{acknowledgments}
		
		This research was funded by the National Research and Development Program of China, No.~2017YFA0303104, the National Natural Science Foundations of China, No.~11774061, and the Shanghai Municipal Science and Technology (Major Project Grant No.~2019SHZDZX01 and No.~20ZR1405300). 
		
	\end{acknowledgments}
		
%
	
	%

\end{document}